\documentclass[12pt]{article}
\usepackage{amsmath,amssymb}
\newcommand{\eqdef}{\stackrel{\text{def}}{=}}
\newcommand{\n}{\nonumber\\}

\newcommand{\ignore}[1]{}
\newcommand{\Romannumeral}[1]{\uppercase\expandafter{\romannumeral#1}}
\newcommand{\I}{\text{\Romannumeral{1}}}
\newcommand{\II}{\text{\Romannumeral{2}}}

\newtheorem{theo}{\bf Theorem}[section]

\allowdisplaybreaks[4]

\begin{document}

\baselineskip=20pt


\newcommand{\Title}[1]{{\baselineskip=26pt
  \begin{center} \Large \bf #1 \\ \ \\ \end{center}}}
\newcommand{\Author}{\begin{center}
  \large \bf  Ryu Sasaki  \end{center}}
\newcommand{\Address}{\begin{center}
    Faculty of Science, Shinshu University,
     Matsumoto 390-8621, Japan\\
      e-mail: ryu@yukawa.kyoto-u.ac.jp
   \end{center}}
\newcommand{\Accepted}[1]{\begin{center}
  {\large \sf #1}\\ \vspace{1mm}{\small \sf Accepted for Publication}
  \end{center}}

\thispagestyle{empty}

\Title{ Symmetric Morse potential 
 is  exactly solvable }

\Author

\Address
\vspace{1cm}

\begin{abstract}
Morse potential $V_M(x)= g^2\exp (2x)-g(2h+1)\exp(x)$ is defined on the full line, $-\infty<x<\infty$
and it defines an exactly solvable 1-d quantum mechanical system with finitely many discrete eigenstates. 
By taking its right half $0\le x<\infty$
and glueing it with the left half of its mirror image $V_M(-x)$, $-\infty<x\le0$, the symmetric Morse potential
$V(x)= g^2\exp (2|x|)-g(2h+1)\exp(|x|)$ is 
obtained. The quantum mechanical system of this  piecewise analytic potential  
 has infinitely many discrete eigenstates with the corresponding eigenfunctions given by the Whittaker W function.
The eigenvalues are the square of the  zeros of the Whittaker function $W_{k,\nu}(x)$ and its linear 
combination with $W'_{k,\nu}(x)$ as a function of $\nu$ with 
fixed $k$ and $x$. This quantum mechanical system seems to offer an interesting example for
discussing the Hilbert-P\'olya conjecture on the pure imaginary zeros of Riemann zeta function $\zeta(s)$ on Re$(s)=\tfrac12$.
\end{abstract}

\subsection*{Keywords:}

piecewise analytic potentials; bound states; Whittaker function;
pure imaginary zeros;
orthogonality theorems; Hilbert-P\'olya conjecture;
associated Hamiltonians;

%

\section{Introduction}
\label{sec:intro}

This is a third paper discussing exact solvability of one dimensional quantum mechanical systems having 
piecewise analytic potentials which are
mirror symmetric $V(-x)=V(x)$ with respect to the origin.
In previous works, a weak attractive piecewise analytic exponential potential 
$V(x)=-g^2\exp(-|x|)$ \cite{sz1} and a confining piecewise analytic exponential potential
$V(x)=g^2\exp(2|x|)$ \cite{sz2,mz2} have been discussed.%
\footnote{The piecewise linear potential $V(x)=|x|$, exactly solvable by
Airy functions \cite{valee}, is probably the first example of this type.}
The present work could be understood as a one parameter generalisation of these results \cite{mz1}.
The eigensystem of the right half of the Morse potential was discussed in detail by Lagarias \cite{lagarias}.
The present paper is a modest supplement of this seminal work.

The spectra or the eigenvalues of this type of exactly solvable quantum mechanical systems are very different from 
those of the `ordinary' exactly solvable systems \cite{infhull}-\cite{os25} based on shape invariance
\cite{genden}, \cite{khare-suh}.
In the latter, the $n$-th  eigenvalue, $E_n$ counted from the ground state, is a simple elementary 
function of $n$; linear, quadratic, inverse quadratic or $q$-quadratic for those belonging to the 
`discrete' quantum mechanics \cite{os24}--\cite{os13}. 
In contrast, the eigenvalues of the Hamiltonian with the symmetric Morse potential are the square of the 
zeros of the
Whittaker function $W_{k,\nu}(2g)$ \cite{whittaker} and its derivative regarded as a function of $\nu$ with fixed $k$ and $g$ \eqref{peven}--\eqref{ne-kappa}.
Correspondingly, the eigenvalues of the piecewise symmetric exponential potential \cite{sz2,mz2} are the 
square of the zeros of the modified 
Bessel function of the second kind $K_{\nu}(g)$ and its derivative regarded as a function of $\nu$ with fixed $g$.
As is well known $K_\nu(x)$ is related to the Whittaker W function \eqref{KWrel}.

With this feature of the spectra, the Hamiltonian system with the symmetric Morse potential 
offers an interesting example for Hilbert-P\'olya conjecture \cite{polya, gasper, lagarias}
on the pure imaginary zeros of Riemann zeta function $\zeta(s)$ on Re$(s)=\tfrac12$.
As for the asymptotic distribution of the zeros (eigenvalues), one can apply WKB approximation or
Bohr-Sommerfeld quantum condition as demonstrated for the symmetric piecewise analytic exponential potential \cite{sz2}.
In this connection we would like to point out the usefulness of certain deformation procedures
applicable to any 1-d quantum mechanical system including the discrete quantum mechanics, in which
the Hamiltonian is a self-adjoint second order difference operator \cite{os24}--\cite{os13}.
By multiple application of Crum's transformations \cite{crum}, one can delete as many lowest lying eigenstates 
as  wanted. By Krein-Adler transformations \cite{krein,adler},  finitely many eigenstates specified  by the 
set $\mathcal{D}=\{d_1,\ldots,d_L\}$ can be deleted so long as the labels satisfy conditions for the positivity of the norm.
Here $d_j\in\mathbb{Z}_{\ge0}$ is the label of the eigenfunction corresponding to the number of nodes.

The present paper is organised as follows. In \S\ref{sec:ori} the essence of the original Morse potential 
on the full line is summarised. Section three is the main part of the paper deriving the eigensystems of the 
symmetric Morse  by imposing matching and finite norm conditions.
In \S\ref{sec:crum} the formulas of the Crum and Krein-Adler transformations are briefly recapitulated.
The corresponding orthogonality relations of the deformed  Hamiltonian systems  
are also presented. The final section is for a summary and comments.

\section{Original Morse Potential }
\label{sec:ori}

Let us recapitulate the main results of the 1-d quantum mechanical system with the original Morse potential 
defined on the full line:
\begin{align}
\mathcal{H}&=-\frac{d^2}{dx^2}+g^2e^{2x}-g(2h+1)e^x,\quad  -\infty<x<\infty,\quad g>0,\ h\in\mathbb{R}.
\label{omorse}
\end{align}
For positive $h>0$, the eigenvalue problem
\begin{align}
\mathcal{H}\phi_n(x)=E_n\phi_n(x),\quad n=0,1,\ldots,
\label{scheq}
\end{align}
has finitely many discrete eigenstates
\begin{align}
  {E}_n&=-(h-n)^2,\quad n=0,1,\ldots,[h]',\\
  \phi_n(x)
  &=\phi_0(x)P_n\bigl(\rho(x)\bigr),\quad
  \phi_0(x)=e^{hx-\tfrac12\rho(x)},\quad \rho(x)\eqdef 2g\exp(x),\\
  P_n\bigl(\rho(x)\bigr)
  &=\bigl(\rho(x)\bigr)^{-n}L_n^{(2h-2n)}\bigl(\rho(x)\bigr),
\end{align}
in which $[n]'$ means the greatest integer not exceeding $n$ and $L_n^{(\alpha)}(x)$ is the Laguerre polynomial of
degree $n$ in $x$. Although  the minimum of the potential exists for $-\tfrac12<h$,
\begin{equation*}
\text{min}V_M(x)=-(h+\tfrac12)^2<0=V_M(-\infty),
\end{equation*}
the system has  continuous spectrum only for  $h\le0$. This is explained by the `zero point energy'.

The system is shape invariant as the potential of the first associated Hamiltonian $\mathcal{H}^{[1]}$ (see \eqref{HLdef} in \S\ref{sec:crum}) 
\begin{equation}
V_M^{[1]}(x)=V_M(x)-2\partial_x^2\log\phi_0(x)=g^2e^{2x}-g(2h-1)e^x,
\label{hmin1}
\end{equation}
has the same form as $V_M(x)$ with $h$ replaced by $h-1$ and $g$ remains unchanged.

\section{Symmetric Morse Potential }
\label{sec:sym}

Now let us discuss the  Schr\"odinger equation \eqref{scheq},
with the symmetric Morse potential
\begin{align}
V(x)&=g^2\exp(2|x|)-g(2h+1)\exp(|x|)=\frac14\rho(x)^2-(h+\tfrac12)\rho(x)\n
&=\frac14\rho(x)^2-k\rho(x),\quad k\eqdef h+\tfrac12, \quad \rho(x)\eqdef 2g\exp(|x|).
\label{symmor}
\end{align}
Here we have introduced parameter $k$ instead of $h$ for convenience and the definition of $\rho(x)$ 
is now mirror symmetric $\rho(-x)=\rho(x)$.
Now the potential grows indefinitely at the boundaries $x=\pm\infty$ and 
 the system has  infinitely many
bound-states with  positive eigenvalues $E_m>0$, on top of the finitely many 
negative eigenvalues $E_m<0$ which could exist when $k>0$.
The
corresponding eigenfunctions must be normalizable, $\psi_m(x) \in
L^2(\mathbb{R})$. Since the potential is parity invariant,
$V(-x)=V(x)$, the eigenfunctions are also parity invariant,
\begin{equation}
\psi_m(-x)=(-1)^m\psi_m(x)\,.
\end{equation}
According to the conventional oscillation theorems \cite{Hille} the
subscript $m$ counts the {\em nodes} in $-\infty<x<\infty$.
Moreover, we may only consider the  positive half-line 
$x\ge0$,
\begin{equation}
\text{even parity}:\quad\psi'_{2n}(0)=0,\qquad \text{odd
parity}:\quad \psi_{2n+1}(0)=0\,,
 \label{bc}
\end{equation}
{\em i.e.,}  with the eigenfunctions constrained by the parity-dependent
boundary condition at the origin.
One could say that 1-d quantum mechanical systems with mirror symmetric potential $V(-x)=V(x)$ 
are equipped two types of eigenfunctions, one satisfying the Neumann boundary condition
at $x=0$ and the other the Dirichlet condition.

\subsection{Eigenfunctions}
\label{sec:eig}

Let us look for the solutions of Schr\"odinger equation \eqref{scheq} 
with the symmetric Morse potential \eqref{symmor} with positive energy $E=\nu^2$, $\nu>0$, in the following form:
\begin{equation}
\psi(x)=\rho(x)^{-\tfrac12}\phi\bigl(\rho(x)\bigr).
\label{psidef}
\end{equation}
It is now rewritten as that for the Whittaker function \cite{whittaker}:
\begin{align}
\text{positive energy:}\quad 
\frac{d^2\phi(\rho)}{d\rho^2}+\left(-\frac14+\frac{k}{\rho}+\frac{\tfrac14+\nu^2}{\rho^2}\right)\phi(\rho)=0.
\label{weq1}
\end{align}
In the same ansatz \eqref{psidef}  the solution with negative (non-positive) 
energy $E=-\mu^2$, $\mu\ge0$ is rewritten as
\begin{align}
\text{negative energy:}\quad 
\frac{d^2\phi(\rho)}{d\rho^2}+\left(-\frac14+\frac{k}{\rho}+\frac{\tfrac14-\mu^2}{\rho^2}\right)\phi(\rho)=0.
\label{weq2}
\end{align}
Among possible sets of general solutions, we choose the following Whittaker W functions.
For the positive energy solutions
\begin{align}
\text{even:}\quad \psi^{(e)}(x)&=\rho(x)^{-\tfrac12}\left(A\, W_{k,i\nu}\bigl(\rho(x)\bigr)+ B\, W_{-k,i\nu}\bigl(-\rho(x)\bigr)\right),
\label{eneg}\\
\text{odd:}\quad \psi^{(o)}(x)&=\rho(x)^{-\tfrac12}\left(C\, W_{k,i\nu}\bigl(\rho(x)\bigr)+ D\, W_{-k,i\nu}\bigl(-\rho(x)\bigr)\right),
\label{oneg}
\end{align}
and for the negative energy solutions
\begin{align}
\text{even:}\quad \psi^{(e)}(x)&=\rho(x)^{-\tfrac12}\left(A\, W_{k,\mu}\bigl(\rho(x)\bigr)+ B\, W_{-k,\mu}\bigl(-\rho(x)\bigr)\right),
\label{epos}\\
\text{odd:}\quad \psi^{(o)}(x)&=\rho(x)^{-\tfrac12}\left(C\, W_{k,\mu}\bigl(\rho(x)\bigr)+ D\, W_{-k,\mu}\bigl(-\rho(x)\bigr)\right).
\label{opos}
\end{align}
The matching condition at the origin \eqref{bc} can be easily met by considering the derivative
\begin{align}
\text{positive energy:}\quad \left.\frac{d\psi^{(e)}(x)}{dx}\right|_{x=0}&=-\frac12\rho_0^{-\tfrac12}\left\{
A\,\bigl(-W_{k,i\nu}(\rho_0)+2\rho_0W'_{k,i\nu}(\rho_0)\bigr)\right.\n
&\left. \hspace{20mm} -B\,\bigl(W_{-k,i\nu}(-\rho_0)+2\rho_0W'_{-k,i\nu}(-\rho_0)\bigr)\right\},\\
\text{negative energy:}\quad \left.\frac{d\psi^{(e)}(x)}{dx}\right|_{x=0}&=-\frac12\rho_0^{-\tfrac12}\left\{
A\,\bigl(-W_{k,\mu}(\rho_0)+2\rho_0W'_{k,\mu}(\rho_0)\bigr)\right.\n
&\left. \hspace{20mm} -B\,\bigl(W_{-k,\mu}(-\rho_0)+2\rho_0W'_{-k,\mu}(-\rho_0)\bigr)\right\},
\end{align}
in which 
\begin{equation*}
\rho_0\eqdef \rho(0)=2g.
\end{equation*}
It is a regular point of Whittaker W functions \eqref{eneg}--\eqref{opos}.
As is clear from the equations \eqref{weq1}, \eqref{weq2}, $\rho=0$ is a regular singular point with
the characteristic exponents $\tfrac12\pm i\nu$ and $\tfrac12\pm \mu$, respectively.
Since $\rho=0$ is not included in the domain of the present problem, another set of solutions
including the Whittaker M functions having these characteristic exponents is irrelevant.

Thus, wave functions satisfying the matching conditions \eqref{bc} at the
origin can be easily found. For positive eigenvalues, they are
\begin{align}
\psi^{(e)}(x)&=\rho(x)^{-\tfrac12}\left(A(k,\nu,\rho_0)\, W_{k,i\nu}\bigl(\rho(x)\bigr)+ B(k,\nu,\rho_0)\, W_{-k,i\nu}\bigl(-\rho(x)\bigr)\right),\n
&\qquad A(k,\nu,\rho_0)\eqdef W_{-k,i\nu}(-\rho_0)+2\rho_0W'_{-k,i\nu}(-\rho_0),\\
&\qquad   B(k,\nu,\rho_0)\eqdef -W_{k,i\nu}(\rho_0)+2\rho_0W'_{k,i\nu}(\rho_0),
\label{pecomb}\\
\psi^{(o)}(x)&=\rho(x)^{-\tfrac12}\left(C(k,\nu,\rho_0)\, W_{k,i\nu}\bigl(\rho(x)\bigr)+ D(k,\nu,\rho_0)\, W_{-k,i\nu}\bigl(-\rho(x)\bigr)\right),\\
&\qquad C(k,\nu,\rho_0)\eqdef -W_{-k,i\nu}(-\rho_0),\quad
              D(k,\nu,\rho_0)\eqdef W_{k,i\nu}(\rho_0).
\label{pocomb}
\end{align}
For negative eigenvalues, they are
\begin{align}
\psi^{(e)}(x)&=\rho(x)^{-\tfrac12}\left(A(k,\mu,\rho_0)\, W_{k,\mu}\bigl(\rho(x)\bigr)+ B(k,\mu,\rho_0)\, W_{-k,\mu}\bigl(-\rho(x)\bigr)\right),\n
&\qquad A(k,\mu,\rho_0)\eqdef W_{-k,\mu}(-\rho_0)+2\rho_0W'_{-k,\mu}(-\rho_0),\\
&\qquad B(k,\mu,\rho_0)\eqdef -W_{k,\mu}(\rho_0)+2\rho_0W'_{k,\mu}(\rho_0),
\label{necomb}\\
\psi^{(o)}(x)&=\rho(x)^{-\tfrac12}\left(C(k,\mu,\rho_0)\, W_{k,\mu}\bigl(\rho(x)\bigr)+ D(k,\mu,\rho_0)\, W_{-k,\mu}\bigl(-\rho(x)\bigr)\right),\\
&\qquad C(k,\mu,\rho_0)\eqdef -W_{-k,\mu}(-\rho_0),\quad
              D(k,\mu,\rho_0)\eqdef W_{k,\mu}(\rho_0).
\label{nocomb}
\end{align}

The  asymptotic condition at $x\to+\infty$, ($\rho\to+\infty$) is easily imposed.
The   Whittaker W function has the following asymptotic behaviour \cite{whittaker} 
\begin{align}
W_{k,\mu}(x)&\sim e^{-\tfrac12 x}\,x^k\left(1+O(\frac1x)\right)\sim W_{k,i\nu}(x), 
\label{Wasym}
\end{align}
as $|x|\to\infty$.
The eigenvalues are selected by requiring 
the coefficients $B(k,\nu,\rho_0)$ \eqref{pecomb} and $D(k,\nu,\rho_0)$ \eqref{pocomb} of the divergent term 
$W_{-k,i\nu}\bigl(-\rho(x)\bigr)$  should vanish for the positive energy eigenstates and 
the coefficients $B(k,\mu,\rho_0)$ \eqref{necomb}
$D(k,\mu,\rho_0)$ \eqref{nocomb} of the divergent term 
$W_{-k,\mu}\bigl(-\rho(x)\bigr)$  should vanish for the negative energy eigenstates.

For $k<0$, the system has  positive energy eigenstates only and they are  numbered by the conditions
\begin{align}
\text{even:}&\quad
-W_{k,i\nu_{2n}}(\rho_0)+2\rho_0W'_{k,i\nu_{2n}}(\rho_0)=0,\quad \quad
n=0,1,\ldots,
\label{peven}\\
\text{odd:}&\qquad \qquad \qquad \qquad \ \  W_{k,i\nu_{2n+1}}(\rho_0)=0,\quad\quad n=0,1,\ldots\,,
\label{podd}
\end{align}
with  the corresponding  eigenfunctions:
\begin{align}
\psi_{2n}(x)&=\rho(x)^{-\tfrac12}W_{k,i\nu_{2n}}\bigl(\rho(x)\bigr),
 \qquad  \qquad   \ \   E_{2n}=\nu_{2n}^2, \quad n=0,1,\ldots,
\label{pevenfun}\\[4pt]
\psi_{2n+1}(x)&=\text{sign}(x)\rho(x)^{-\tfrac12}W_{k,i\nu_{2n+1}}\bigl(\rho(x)\bigr),
\ \  E_{2n+1}=\nu_{2n+1}^2, \quad n=0,1,\ldots,
\label{poddfun}\\
&  0<g(g-k)<E_0<E_1<E_2<\cdots \  \Leftrightarrow \
\sqrt{g(g-k)}<\nu_0<\nu_1<\nu_2<\cdots . 
\label{pe-kappa}
\end{align}
For $k>0$, there are approximately $k-1$ eigenstates with negative energy. 
These eigenvalues are determined by the conditions
\begin{align}
\text{even:}&\quad
-W_{k,\mu_{2n}}(\rho_0)+2\rho_0W'_{k,\mu_{2n}}(\rho_0)=0,\quad \quad
n=0,1,\ldots,
\label{oeven}\\
\text{odd:}&\qquad \qquad \qquad \qquad \ \  W_{k,\mu_{2n+1}}(\rho_0)=0,\quad\quad n=0,1,\ldots\,.
\label{nodd}
\end{align}
The corresponding  eigenfunctions are
\begin{align}
\psi_{2n}(x)&=\rho(x)^{-\tfrac12}W_{k,\mu_{2n}}\bigl(\rho(x)\bigr),
 \qquad  \qquad   \ \   E_{2n}=-\mu_{2n}^2, \quad n=0,1,\ldots,
\label{nevenfun}\\[4pt]
\psi_{2n+1}(x)&=\text{sign}(x)\rho(x)^{-\tfrac12}W_{k,\mu_{2n+1}}\bigl(\rho(x)\bigr),
\ \  E_{2n+1}=-\mu_{2n+1}^2, \quad n=0,1,\ldots,.
\label{noddfun}\\
&  -k^2<E_0<E_1<E_2<\cdots <0 \ \Leftrightarrow \ 
k>\mu_0>\mu_1>\mu_2>\cdots >0. 
\label{ne-kappa}
\end{align}
The eigenstates with positive eigenvalues are numbered after the negative ones.
The eigenfunctions have the same form as \eqref{pevenfun}, \eqref{poddfun} and the eigenvalues are
 determined by the same conditions \eqref{peven} and \eqref{podd} but the numbering follows that 
of the negative energy ones.

For $k=0$ the symmetric Morse potential \eqref{symmor} reduces to the confining piecewise analytic
exponential potential $V(x)=g^2\exp(2|x|)$ discussed in a previous paper \cite{sz2}.
 It has positive energy eigenvalues only and its eigenfunctions are
the modified Bessel function of the second kind $K_{i\nu}(x)$, which is related to Whittaker W function
(\cite{erdelyi} Vol. 1, \S6.9 formula (14))
\begin{equation}
K_{\alpha}(x)=\sqrt{\frac{\pi}{2x}}\,W_{0,\alpha}(2x).
\label{KWrel}
\end{equation}
The factor 2 among the arguments is reflected by the factor two in the definitions of $\rho(x)$ in \cite{sz2}
and in this paper. This also explains the extra factor $\rho(x)^{-\tfrac12}$ in the wavefunction $\psi(x)$  formula \eqref{psidef}
compared to the counterpart in \cite{sz2}.
By using \eqref{KWrel} one can deduce the condition \eqref{peven} gives
$K'_{i\nu_{2n}}(g)=0$ when $k=0$.

In this manner  the exact solvability of the symmetric Morse potential \eqref{symmor} is established.
The orthogonality relations among the eigenfunctions have the following forms:
\begin{align}
\int_0^{\infty}e^{-x}W_{k,i\nu_{2n}}(2g\,e^{x})W_{k,i\nu_{2m}}(2g\,e^{x})\,dx&
\propto\delta_{n\,m},
\label{porteven}\\
\int_0^{\infty}e^{-x}W_{k,i\nu_{2n+1}}(2g\,e^{x})W_{k,i\nu_{2m+1}}(2g\,e^{x})\,dx&
\propto\delta_{n\,m},
 \label{portodd}\\
 \int_0^{\infty}e^{-x}W_{k,\mu_{2n}}(2g\,e^{x})W_{k,\mu_{2m}}(2g\,e^{x})\,dx&
\propto\delta_{n\,m},
\label{norteven}\\
\int_0^{\infty}e^{-x}W_{k,\mu_{2n+1}}(2g\,e^{x})W_{k,\mu_{2m+1}}(2g\,e^{x})\,dx&
\propto\delta_{n\,m},
 \label{nortodd}\\
\int_0^{\infty}e^{-x}W_{k,\mu_{2n}}(2g\,e^{x})W_{k,i\nu_{2m}}(2g\,e^{x})\,dx&
\propto\delta_{n\,m},
\label{nporteven}\\
\int_0^{\infty}e^{-x}W_{k,\mu_{2n+1}}(2g\,e^{x})W_{k,i\nu_{2m+1}}(2g\,e^{x})\,dx&
\propto\delta_{n\,m}.
 \label{nportodd}
\end{align}

%
%
\subsection{Deformed Hamiltonians}
\label{sec:crum}

When a 1-d Hamiltonian (Sturm-Liouville) system $\{\mathcal{H},E_n,\psi_n\}$ is given,
it is possible to construct  deformed systems in which finitely many eigenvalues and corresponding eigenfunctions are
deleted. The simplest one due to Crum \cite{crum}
 is  to delete the lowest lying $L$ levels $\{E_j,\psi_j(x)\}$, $j=0,1,\ldots,L-1$
from the the original one-dimensional Hamiltonian system
$\mathcal{H}=\mathcal{H}^{[0]}$, 
$\{E_n,\psi_n(x)\}$, $n=0,1,\ldots$. The deformed  Hamiltonian systems $\mathcal{H}^{[L]}$ $L=1,2,\ldots$,
are {\em essentially
iso-spectral}, that is, the remaining eigenvalues are unchanged:
\begin{align}
\mathcal{H}^{[L]}\psi_n^{[L]}(x)&=E_n\psi_n^{[L]}(x),
\quad n=L,L+1,\ldots,\\
\mathcal{H}^{[L]}&\eqdef\mathcal{H}^{[0]}
-2\partial_x^2\log\left|\text{W}[\psi_0,\psi_1,\ldots,\psi_{L-1}](x)\right|,
\label{HLdef}\\
\psi_n^{[L]}(x)&\eqdef\frac{\text{W}[\psi_0,\psi_1,\ldots,
\psi_{L-1},\psi_n](x)}{\text{W}[\psi_0,\psi_1,\ldots,\psi_{L-1}](x)},\quad
(\psi_n^{[L]},\psi_m^{[L]})=\prod_{j=0}^{L-1}(E_n-E_j)(\psi_n,\psi_m),
\label{wronM}
\end{align}
in which the Wronskian of $n$-functions $\{f_1,\ldots,f_n\}$ is
defined by formula
\begin{align}
&\text{W}\,[f_1,\ldots,f_n](x)
  \eqdef\det\Bigl(\frac{d^{j-1}f_k(x)}{dx^{j-1}}\Bigr)_{1\leq j,k\leq n}.
  \label{wron}
  \end{align}
This result is obtained from a multiple application of the Darboux
transformations. 

Another deformation method is due to Krein \cite{krein} and Adler \cite{adler}.  
It deletes finitely many eigenlevels specified by the set 
$\mathcal{D}=\{d_1,d_2,\ldots,d_L\}$, $d_j\in\mathbb{Z}_{\ge0}$ satisfying the conditions
\begin{equation}
\prod_{j=1}^L(m-d_j)\ge0,\quad \forall m\in\mathbb{Z}_{\ge0}.
\label{poscon}
\end{equation}
The deformed Hamiltonian system $\{\mathcal{H}_{\mathcal D}, E_n,\psi_{\mathcal{D};n}(x)\}$,
is given by 
\begin{align}
\mathcal{H}_{\mathcal D}\psi_{\mathcal{D};n}(x)&=E_n\psi_{\mathcal{D};n}(x),
\quad n\in\mathbb{Z}_{\ge0}\backslash\mathcal{D},\\
\mathcal{H}_{\mathcal D}&\eqdef\mathcal{H}^{[0]}
-2\partial_x^2\log\left|\text{W}[\psi_{d_1},\psi_{d_2},\ldots,\psi_{d_L}](x)\right|,\\
\psi_{\mathcal{D};n}(x)&\eqdef\frac{\text{W}[\psi_{d_1},\psi_{d_2},\ldots,
\psi_{d_L},\psi_n](x)}{\text{W}[\psi_{d_1},\psi_{d_2},\ldots,\psi_{d_L}](x)},\quad
(\psi_{\mathcal{D};n},\psi_{\mathcal{D};m})=\prod_{j=1}^{L}(E_n-E_{d_j})(\psi_n,\psi_m).
\label{wronKA}
\end{align}
The above conditions  on the set $\mathcal{D}$ \eqref{poscon} are necessary and sufficient for the positivity of norms and self-adjointness 
of the deformed Hamiltonian $\mathcal{H}_{\mathcal D}$.

Let us apply Crum's sequence to the present Hamiltonian \eqref{scheq}, \eqref{symmor}, 
\eqref{pevenfun}--\eqref{pe-kappa}, \eqref{nevenfun}--\eqref{ne-kappa}.
Parallel expressions for the Krein-Adler deformations can be obtained easily.
 It is easy to see that the systems are
parity invariant:
\begin{align}
V^{[L]}(x)&\eqdef V(x)-2\partial_x^2\log
\left|\text{W}[\psi_0,\psi_1,\ldots,\psi_{L-1}](x)\right|,
\quad V^{[L]}(-x)=V^{[L]}(x),\\
\psi_n^{[L]}(-x)&=(-1)^{L+n}\psi_n^{[L]}(x).
\end{align}
Because of the parity, the orthogonality relations among the even
and odd eigenfunctions are trivial and those even-even and odd-odd
\begin{align}
\delta_{n\,m}\propto
(\psi_n^{[L]},\psi_m^{[L]})=\int_{-\infty}^{\infty}
\psi_n^{[L]}(x)\psi_m^{[L]}(x)dx
\end{align}
can be rewritten as those on the positive $x$-axis
\begin{align}
\delta_{n\,m}&\propto \int_{0}^{\infty}
\psi_{2n}^{[L]}(x)\psi_{2m}^{[L]}(x)dx,
\label{ortMeven}\\
\delta_{n\,m}&\propto \int_{0}^{\infty}
\psi_{2n+1}^{[L]}(x)\psi_{2m+1}^{[L]}(x)dx. \label{ortModd}
\end{align}

In the following we consider the case of $k\le0$ so that all the eigenvalues 
are positive. For the case $k>0$, similar expressions can be obtained with 
relatively more notational complication.
By using known properties of the Wronskians \cite{sz1},
we can reduce the Wronskians
of $\{\psi_n(x)\}$ in  \eqref{wronM}  to the Wronskians of the
Whittaker W function  $\{W_{k,i\nu_n}(\rho)\}$. 
This makes the actual evaluation much simpler, for
example, we obtain for $x>0$:
\begin{align}
\text{W}[\psi_0,\psi_n](x)&=
\text{W}[W_{k,i\nu_0}(\rho),W_{k,i\nu_n}(\rho)](\rho),\n[-6pt]
\cdots\qquad & \qquad \cdots\n
\text{W}[\psi_0,\psi_1,\ldots,\psi_{L-1},\psi_n](x)&=
\rho^{(L-1)(L+1)/2}\cdot\n[-6pt]
&\quad\times
\text{W}[W_{k,i\nu_0}(\rho),\ldots,W_{k,i\nu_{L-1}}(\rho),
W_{k,i\nu_n}(\rho)](\rho).
\end{align}

It is straightforward to evaluate $V^{[L]}(x)$ asymptotically  by using that of Whittaker W function
$W_{k,\nu}(x)$ \eqref{Wasym}: It has the form 
\begin{align*}
V^{[L]}(x)=\frac14\rho(x)^2-(k-L)\rho(x)+O(\frac1x),\qquad |x|\to\infty,
\end{align*}
which is {\em not shape invariant} but the parameter $k (h)$ retains the property of the original
Morse potential \eqref{hmin1}.

The results obtained in the previous subsection can be stated as 
various Theorems on Whittaker W functions:

\begin{theo}{\bf Pure imaginary zeros}\label{theo:one}\quad
When Whittaker W functions $W_{k,\nu}(x)$, $\frac{d}{dx}W_{k,\nu}(x)$
are regarded as functions of the parameter $\alpha$ for fixed $k$ and $x>0$,
they have infinitely many pure imaginary zeros:
\begin{align} 
-W_{k,i\lambda_j}(x)+2x\frac{dW_{k,i\lambda_j}(x)}{dx}&=0,
\qquad 0<\frac{x}2<\lambda_0<\lambda_1<\lambda_2<\ \cdots, 
   \label{lams}\\
   W_{k,i\eta_j}(x)&=0, \quad 0<\frac{x}2<\eta_0<\eta_1<\eta_2<\ \cdots.
   \label{etas}
\end{align}
They are interlaced by the oscillation theorem:
\begin{equation}
0<\frac{x}2<\lambda_0<\eta_0<\lambda_1<\eta_1<\lambda_2<\eta_2<\,\cdots .
\end{equation}
\end{theo}
Since the discrete eigenvalues of one dimensional quantum mechanics are always simple, 
all these zeros are also simple.
\begin{theo}{\bf Orthogonality relation \I}\label{theo:two}\quad 
The Whittaker W function  with the above pure imaginary parameters $\{i\lambda_j\}$ \eqref{lams},
$\{i\eta_j\}$ \eqref{etas} satisfy the following orthogonality relations \rm{(}$x>0$\rm{)}:
\begin{align}
\text{even}:& \quad
\int_x^{\infty}W_{k,i\lambda_j}(\rho)W_{k,i\lambda_k}(\rho)\frac{d\rho}{\rho^2}=0,\qquad
j\neq k,
\label{theo1even}\\
\text{odd}:& \quad
\int_x^{\infty}W_{k,i\eta_j}(\rho)W_{k,i\eta_j}(\rho)\frac{d\rho}{\rho^2}=0,\qquad
j\neq k. \label{theo1odd}
\end{align}
\end{theo}
Let us denote these two types of zeros by one consecutive sequence
($\{\nu_j\}$):
\begin{align*}
\nu_0\equiv \lambda_0,\ \nu_1\equiv \eta_0,\ \nu_2\equiv \lambda_1,\
\nu_3\equiv \eta_1,\ldots,.
\end{align*}

The  orthogonality relations of the eigenfunctions
\eqref{ortMeven}--\eqref{ortModd} of the $L$-th associated
Hamiltonian system can be stated as\
\begin{theo}{\bf Orthogonality relation \II}\label{theo:thr}
\begin{align}
&{\rm even}:\int_x^{\infty}\!\frac{\text{W}[W_{k,i\nu_0},\ldots,W_{k,i\nu_{L-1}},
W_{k,i\nu_{2n}}](\rho)\text{W}[W_{k,i\nu_0},\ldots,W_{k,i\nu_{L-1}},
W_{k,i\nu_{2m}}](\rho)}
{\left(\text{W}[W_{k,i\nu_0},\ldots,W_{k,i\nu_{L-1}}](\rho)\right)^2}\rho^{2(L-1)}
d\rho=0,\n
& \hspace{135mm} n\neq m,
\label{theo2even}\\
&{\rm odd}:\!\!\int_x^{\infty}\!\!\frac{\text{W}[W_{k,i\nu_0},\ldots,\!W_{k,i\nu_{L-1}},
\!W_{k,i\nu_{2n+1}}](\rho)\!\text{W}[W_{k,i\nu_0},\ldots,\!W_{k,i\nu_{L-1}},
\!W_{k,i\nu_{2m+1}}](\rho)}
{\left(\text{W}[W_{k,i\nu_0},\ldots,W_{k,i\nu_{L-1}}](\rho)\right)^2}\rho^{2(L-1)}
d\rho=0,\n 
&\hspace{135mm}
n\neq m. \label{theo2odd}
\end{align}
\end{theo}
Theorem\,\ref{theo:two} is the special case ($L=0$) of Theorem\,\ref{theo:thr}.

As for the asymptotic distribution of the pure imaginary zeros $\{\nu_n\}$, $n\gg1$ \cite{lagarias}, 
we can make a conjecture based on the WKB approximation or the so-called Bohr-Sommerfeld 
quantum condition
$\oint p(x)dx=2\pi (n+\tfrac12)$. Here $p(x)$ is the momentum at $x$  determined by the
energy conservation $p(x)^2+g^2\,e^{2x}-2gke^x=E_n=\nu_n^2$. In terms of the elementary integral 
\begin{equation}
4\int_0^{\log[(k+\sqrt{k^2+\nu_n^2})/g]}\!\!\sqrt{\nu_n^2-g^2\,e^{2x}+2gke^x}\,dx=2\pi (n+\tfrac12),
\end{equation}
the asymptotic dependence of $\nu_n$ on $n$ is obtained.

\section{Summary and Comments}
\label{summary}

Following the examples
of the weak attractive piecewise analytic exponential potential
$V(x)=-g^2\exp(-|x|)$ \cite{sz1}, the confining non-analytic exponential potential
$V(x)=g^2\exp(2|x|)$ \cite{sz2,mz2} and the half line Morse potential \cite{lagarias}, 
the exact solvability of the quantum system with the 
symmetric Morse potential \eqref{symmor} is demonstrated. 
Certain similarity to the original Morse potential is observed.
Depending on the sign of the parameter $k$, the system has positive energy eigenstates only ($k\le0$)
and positive and finitely many negative eigenvalues ($k>0$). 
The mirror symmetric potential imposes the Neumann boundary condition for the even level
eigenfunctions and the Dirichlet for the odd level eigenfunctions.
The eigenvalues are determined as the 
zeros of the Whittaker W function $W_{k,\nu}(x)$ and its linear combination with  $W'_{k,\nu}(x)$
regarded as the function of $\nu$  with fixed $k$ and $x$.
Resulting orthogonality relations among the eigenfunctions are explored in some detail.
Two types of deformed Hamiltonian systems are mentioned for possible relevance to 
Hilbert-P\'olya conjecture. These deformations could be used to enhance the precision 
of numerical fitting of the eigenvalues of any model by allowing to delete finitely many eigenvalues
subject to certain conditions.

For possible relevance to Hilbert-P\'olya conjecture, it would be desirable to generate many more
quantum mechanical Hamiltonian systems having similar features to the present example, hopefully
with more parameters. One direction would be to look for systems having (confluent) basic
hypergeometric ($q$-hypergeometric) functions as eigenfunctions. 
The `discrete' quantum mechanics \cite{os24,os12,os13} have many such examples.

\subsection*{Acknowledgements}

The author thanks Jeffrey Lagarias for enlightening communication.
He also thanks Milosh Znojil for many interesting discussions on exact solvability.




\end{document}